\documentclass[prd,amsmath,notitlepage,onecolumn,12pt,nofootinbib]{revtex4-2}
\usepackage{graphicx}
\usepackage{float}
\usepackage{epstopdf,cancel}
\usepackage{epsf,latexsym,bbm,euscript}
\usepackage{amssymb,amsmath}
\usepackage{mathtools}
\usepackage{times,graphics}
\usepackage{soul,xcolor}
\usepackage{mathtools}
\usepackage[normalem]{ulem}
\usepackage[margin=1in]{geometry}
\usepackage{pifont}
\usepackage[utf8]{inputenc}
\usepackage{tensor}
\usepackage[utf8]{inputenc}
\usepackage{graphicx}
\usepackage{float}
\usepackage{epsf,latexsym,bbm,euscript}
\usepackage{amssymb,amsmath}
\usepackage{mathtools}
\usepackage{times}
\usepackage{soul,xcolor}
\usepackage{mathtools}
\usepackage{mathrsfs}
\usepackage{array}
\usepackage{booktabs}
\usepackage{enumitem}

\usepackage{xcolor}

\newcommand{\pbh}{$\Phi$BH}

\usepackage[caption=false]{subfig}
\newcommand{\be}{\begin{equation}}
	\newcommand{\ee}{\end{equation}}

\def\6{{\langle}}
\def\9{{\rangle}}
\def\pad{{\partial}}

\newcommand{\defeq}{\vcentcolon=}

\usepackage{scalerel}
\usepackage{tikz}
\usetikzlibrary{svg.path}
\definecolor{orcidlogocol}{HTML}{A6CE39}
\tikzset{
	orcidlogo/.pic={
		\fill[orcidlogocol] svg{M256,128c0,70.7-57.3,128-128,128C57.3,256,0,198.7,0,128C0,57.3,57.3,0,128,0C198.7,0,256,57.3,256,128z};
		\fill[white] svg{M86.3,186.2H70.9V79.1h15.4v48.4V186.2z}
		svg{M108.9,79.1h41.6c39.6,0,57,28.3,57,53.6c0,27.5-21.5,53.6-56.8,53.6h-41.8V79.1z M124.3,172.4h24.5c34.9,0,42.9-26.5,42.9-39.7c0-21.5-13.7-39.7-43.7-39.7h-23.7V172.4z}
		svg{M88.7,56.8c0,5.5-4.5,10.1-10.1,10.1c-5.6,0-10.1-4.6-10.1-10.1c0-5.6,4.5-10.1,10.1-10.1C84.2,46.7,88.7,51.3,88.7,56.8z};
	}
}
\newcommand\orcidlink[1]{\href{https://orcid.org/#1}{\mbox{\scalerel*{
				\begin{tikzpicture}[yscale=-1,transform shape]
					\pic{orcidlogo};
				\end{tikzpicture}
			}{X}}}}

\def\sg{\textsl{g}}

\newcommand{\cmark}{\text{\ding{51}}}
\newcommand{\xmark}{\text{\ding{55}}}

\usepackage{url,hyperref}
\hypersetup{colorlinks,linkcolor={blue!55!black},citecolor={red!45!black},urlcolor={blue!45!black},breaklinks=true}
\begin{document}

	\title{Exotic Encounters: Buchdahl’s Conditions and Physical Black Holes}
	\author{Ioannis Soranidis,\orcidlink{0000-0002-8652-9874}}
	\email{ioannis.soranidis@mq.edu.au}
	\author{Daniel R. Terno,\orcidlink{0000-0002-0779-0100}}
	\thanks{Corresponding author: \href{mailto:daniel.terno@mq.edu.au}{daniel.terno@mq.edu.au}}
	\affiliation{School of Mathematical and Physical Sciences, Macquarie University, Sydney, New South Wales 2109, Australia}
	\setcounter{page}{0}
	
	\begin{abstract}
		\vspace*{1cm}
Black holes are among the most well-known astrophysical objects, yet their physical realisation remains conceptually subtle. We analyse physical black holes --- light-trapping regions that form in finite time as seen by a distant observer --- and investigate the properties of the matter required to support them. Taking Buchdahl’s theorem as a benchmark, we show that these configurations necessarily violate at least two of its four original conditions, and the post-formation state violates them all. These violations are substantial: they include the null energy condition, non-monotonic energy profiles, and strong pressure anisotropies. Thus, the requirement of truly forming a horizon places physical black holes in a class of solutions that are more exotic than exotic compact objects.

	\end{abstract}

	\maketitle
	
	\vspace{20mm}
	\begin{center}
	\textit{Awarded an Honorable Mention for the Gravity Research Foundation 2025 Awards\\ for Essays on Gravitation.}
	\end{center}

	\vspace{15mm}
	
%	\text{\hfill \hspace*{104mm} Submitted on March 28, 2025}
	
	\vspace{5mm}

	\newpage
	
	\section{Introduction}\label{sec:introduction}
	
	The zoo of exotic compact objects \cite{BCNS:19,CP:19} is perhaps best entered through a 1959 paper by Buchdahl \cite{B:59}. It posed the question: under the rules of general relativity, how compact can a static, spherical object be before collapsing into what we now call a black hole? Given four apparently reasonable assumptions, its radius must satisfy $r_0 \geqslant 9 r_\sg /8 = 9 M/4$, where $r_\sg$ and $M$ are the Schwarzschild radius of the object and its mass, respectively.

Always known to those who know things well, Buchdahl’s result remained in the background. When the three classics --- Weinberg \cite{w:book:72}, MTW \cite{MTW:73}, and Hawking–Ellis \cite{HE:73} --- were published, black holes had already got their name and their first astrophysical candidate, Cygnus X-1. The inevitability of collapse was discussed at length, but Ref.~\cite{B:59} was not cited there, nor in either Wald \cite{W:84} or Chandrasekhar \cite{C:92} a decade later. By the time it was absent in the monumental \textit{Black Hole Physics} of Frolov and Novikov \cite{FN:98}, the first precise measurements of pulsar binaries were already being made. In the new millennium, motion of the stars around Sagittarius A$^*$ was precisely tracked, and \textit{The Relativist’s Toolkit} by Poisson \cite{P:04} did not mention Ref.~\cite{B:59}. The Buchdahl limit was an important but niche topic, discussed, e.g., in \textit{Relativistic Hydrodynamics} \cite{RZ:13} or the encyclopedic \textit{Exact Solutions} \cite{exact:03}.

Things have changed in the last ten years. We have heard the chirp of colliding astrophysical black holes (ABHs) and seen them as shadows in the sky --- all thanks to the amazing advances in interferometry and computer simulations \cite{GEG:24}. More than a hundred dark, ultracompact, and massive objects have already been identified, and the advent of multi-messenger astronomy is expected to uncover many more ABHs \cite{BCNS:19,CP:19,LISA:22}.

	Beyond astronomers studying black holes, theoretical physicists are equally intrigued by them, though for largely different reasons. Black holes provide access to the strong gravity regime and may offer insights into dark matter, the fate of spacetime singularities, or the effects of quantum gravity \cite{BCNS:19,LISA:22}. However, so far ABHs are the only evidence that black holes, understood as objects with horizon(s), are really present in our observable universe. Moreover, assuming that ABHs have horizons comes with a conceptual price \cite{CP:19,MMT:22}. The exteriors of Schwarzschild or Kerr black holes are regular, but their interiors  are not. They contain Cauchy horizons and singularities. Such pathologies are expected to be resolved by a presently unknown quantum theory of gravity, but the known quantum effects bring a host of technical difficulties and unresolved paradoxes, the most celebrated case being the infamous information loss problem.
	
	All of the above motivate alternative views. They postulate  the existence of some black hole mimickers that fit the observed data (and are thus sufficiently close to the classic solutions of general relativity), but are pathology-free \cite{BCNS:19,CP:19,LISA:22,MMT:22}. Beyond regular black holes --- light-trapping domains that are, nevertheless, singularity-free --- there is a variety of hypothetical objects that lack a horizon  and are thus labeled as horizonless exotic compact objects (ECOs).
	
	To quantify in a rigorous way how close a self-gravitating body fits into these categories, it is customary to introduce the so-called closeness parameter, which in spherical symmetry is naturally defined as $\epsilon \defeq1- 2M/r_0$ \cite{CP:19}. Black holes, independent of whether they are regular or not, are characterized by $\epsilon = 0$, while horizonless configurations are described by $\epsilon > 0$. The Buchdahl limit sets $\epsilon \geqslant \tfrac{1}{9}$, and if one wishes to model a more compact object, some of the assumptions --- either directly or via hypothetical quantum effects or modifications of general relativity --- should be violated. Thus, the assumptions of Ref.~\cite{B:59} serve as model classifiers: their violations typically require ``exotic" matter.

This is how ECOs got their ``exotic" appellation. We show that if black hole horizons are observable, then they are even more exotic. They violate all the assumptions behind the Buchdahl limit, typically simultaneously. What to do with this, is another matter.
	
	\section{Collapse and its limits}

We begin with the assumptions underlying Ref.~\cite{B:59}, and their modern sharpening in Ref.~\cite{A:08}. Beyond the validity of general relativity and spherical symmetry, there are two common assumptions:
\begin{enumerate}
    \renewcommand{\labelenumi}{{\Roman{enumi}}:}
    \item Energy density $\rho$ and radial pressure $p$ are non-negative, i.e., $\rho \geqslant 0$ and $p \geqslant 0$
    \item The object is described by a single perfect fluid
\end{enumerate}
Two additional assumptions of Ref.~\cite{B:59} are:
\begin{enumerate}
    \renewcommand{\labelenumi}{B\textsubscript{\arabic{enumi}}:}
    \item The fluid may be mildly anisotropic, with the tangential pressure $p_\|$ satisfying $p \geqslant p_\|$
    \item Decreasing energy density profile, i.e., $\rho'(r) < 0$
\end{enumerate}

We note that a non-static, anisotropic fluid ($p \neq p_\|$) requires at least two ideal fluids to reproduce the energy-momentum tensor (EMT) \cite{RZ:13}. The last two assumptions are replaced in Ref.~\cite{A:08} by:
\begin{enumerate}
    \renewcommand{\labelenumi}{A:}
    \item The anisotropy is bounded by $p + 2p_\| \leqslant K\rho$, for some $K > 0$.
\end{enumerate}
As a result, the bound on a static configuration becomes
\be
\epsilon \geqslant (1 + 2K)^{-2},
\ee
where $K = 1$ recovers the Buchdahl limit and any model which satisfies the dominant energy condition has $K \leqslant 3$.

From the point of view of a distant observer, once such a bound is broken, there are three possible outcomes \cite{MMT:22,M:23}:
\begin{enumerate}[label=(\roman*)]
    \item Perpetual ongoing collapse ($\epsilon \rightarrow 0$ as $t \rightarrow \infty$): the horizon exists only as an asymptotic ($t \rightarrow \infty$) concept. At times $t < \infty$, these frozen stars remain horizonless by definition.
    \item Formation of an ECO: the closeness parameter reaches a minimal value $\epsilon_{\min} > 0$, either at some finite time $t_{\min}$ or asymptotically as $t \rightarrow \infty$.
    \item Formation of an apparent horizon in finite time $t_{\mathrm{f}} < \infty$, as measured by the clock of a distant observer.
\end{enumerate}

The Kerr and Schwarzschild solutions are the asymptotic limit of (i). ECOs by definition belong to (ii). If we are interested in black holes  that have formed by now, we have to select (iii).\footnote{We note that the collapse process can also be described through a virial equilibrium approach, in which the system first reaches the Buchdahl limit. In this context, it is important to emphasize that the Buchdahl theorem involves matching an interior solution to an exterior vacuum, spherically symmetric geometry described by the Schwarzschild metric. Only if this intermediate configuration is unstable does the collapse proceed to form a black hole. For further details on this perspective, we refer the reader to Refs.~\cite{DG:24,D:25}. In contrast, our approach explicitly assumes that the final state of the collapse is a (dynamical) black hole has been reached, and the formalism developed in the following sections is specifically designed to extract necessary conditions for existence of such configurations.}

There is a large discrepancy in what different disciplines mean by ``black hole'' \cite{C:19}. ABHs are the ultracompact objects that have been observed. Following Frolov, we consider two more classes of black holes \cite{F:14}. \textit{Mathematical} black holes are entities with one or more event horizons, such as Kerr solutions or various static regular black holes. Event horizons are inherently teleological and fundamentally unobservable \cite{HE:73,FN:98,F:14}. Their determination requires knowledge of the entire history of the universe, and even the infalling observer’s experiences --- and whatever adventures await beyond the horizon --- are counterfactual, since they cannot be observed or communicated. A locally defined apparent horizon, which is the boundary of a light-trapping region, is what is actually determined in numerical simulations, particularly in dynamical scenarios. In general, it also depends on the foliations of spacetime, which are observer-dependent.

In contrast with mathematical black holes, a \textit{physical} black hole (\pbh\footnote{We normally follow Ref.~\cite{F:14} and use the abbreviation PBH. However, since this is also the standard acronym for primordial black hole, we adopt a distinct abbreviation here to avoid confusion.}) is a light-trapping spacetime domain that we additionally require to form within a finite time as measured by a distant observer \cite{MMT:22,M:23}.

\section{Exotic nature of Physical black holes}

Our self-consistent analysis is applicable to any theory that is based on a metric description of gravity. The metric $\sg_{\mu\nu}$ is a solution of the Einstein equations
\be
G_{\mu\nu}= T_{\mu\nu}\equiv 8\pi\6\hat{T}_{\mu\nu}\9_\omega, \label{ee}
\ee
where the left-hand side is the Einstein tensor $G_{\mu\nu} = R_{\mu\nu} - \tfrac{1}{2}R\sg_{\mu\nu}$, and we make no specific assumptions about the EMT on the right. It includes the renormalised expectation value of all matter fields, higher-order terms arising from its regularisation, and possible contributions arising from modifications to Einstein--Hilbert action or a cosmological constant $\Lambda$. Our analysis does not use any specific property of the state $\omega$ and does not separate the matter EMT into the collapsing matter and (perturbatively-obtained) quantum excitations.

We focus on spherical symmetry (and, to simplify exposition, on asymptotically flat spacetimes, though this is not essential \cite{DMSST:24}). Then any metric in Schwarzschild coordinates $(t,r)$ has the form
\begin{align}
ds^2 = -e^{2h(t,r)}f(t,r)dt^2 + f(t,r)^{-1}dr^2 + r^2d\Omega_2\ ,
\end{align}
where $r$ is the areal radius and $t$ is the proper time of a distant static observer.

The function $f$ is coordinate-independent and is conveniently expressed via the Misner--Sharp--Hernandez (MSH) mass as  \cite{RZ:13,MMT:22}:
\be
f = 1 - \frac{2M(t,r)}{r} \defeq \pad_\mu r \pad^\mu r\ .
\ee
The functions $h$ and $h_+$ play the role of integrating factors in the coordinate transformations, such as
\be
dt = e^{-h}(e^{h_+}dv - f^{-1}dr)\ , \label{trvr-transformation}
\ee
where $v$ is the advanced null coordinate.

The outer boundary of the trapped (or antitrapped) region is given by the largest root $r_\sg(t)$ of $f = 0$, i.e., $r_\sg = 2M(t,r_\sg)$. We impose only the weakest regularity requirement: there are no divergent curvature scalars \cite{exact:03} up to and on the apparent horizon at $r = r_\sg(t)$.

Both the Einstein equations and the regularity conditions simplify considerably if we introduce
\begin{align}
\tau{_t} \defeq e^{-2h} {T}_{tt} \equiv \rho f\ , \qquad {\tau}{^r} \defeq T^{rr} \equiv p f\ , \qquad \tensor{\tau}{_t^r} \defeq e^{-h} \tensor{T}{_t^r} \equiv -\phi f\ , \label{eq:mtgEMTdecomp}
\end{align}
where $\rho$, $p$, and $\phi$ are the energy density, radial pressure, and (outgoing) flux, respectively, as measured in an orthonormal frame constructed from the Schwarzschild coordinates (hereafter: Schwarzschild frame). Spherically symmetric spacetimes are described by four independent Einstein equations. Two of the four serve to enforce conservation of the EMT, so of the three equations
\begin{align}
\partial_r C = 8\pi r^2 {\tau}{_t} / f\ , \qquad \partial_t C = 8\pi r^2 e^h \tensor{\tau}{_t^r}\ , \qquad \partial_r h = 4\pi r ({\tau}{_t} + {\tau}{^r}) / f^2\ , \label{eq:Einstein-C}
\end{align}
one is a constraint among the parameters of the metric functions (or the EMT) that appear in a formal series solution of the two others.

It is sufficient to enforce convergence of only two scalars, namely $R$ and $R_{\mu\nu}R^{\mu\nu}$, as $r \to r_\sg$, in order to ensure regularity of all other curvature scalars \cite{MMT:22}. Moreover, it is enough to ensure finite values of the limits of the parts obtained from the $(t,r)$ blocks,
\be
\mathrm{T} = ({\tau}{^r} - {\tau}{_t}) / f\ , \qquad \mathfrak{T} = \big(({\tau}{^r})^2 + ({\tau}{_t})^2 - 2(\tensor{\tau}{_t^r})^2\big) / f^2\ ,
\ee
respectively. The requirement of absence of curvature singularities and existence of real-valued solutions restricts the possible scalings of the EMT components near the apparent horizon to
\begin{align}
\tau_t \propto f^k, \qquad \tau^r \propto f^k, \qquad \tensor{\tau}{_t^r} \propto f^k,
\end{align}
with $k = 0$ or $k = 1$.

Solutions with $k = 0$ satisfy
\be
\tau_t \to -\Upsilon^2(t), \quad \tau^r \to -\Upsilon^2(t), \qquad \tensor{\tau}{_t^r} \to \mp \Upsilon^2(t),
\ee
with the metric functions given by
\be
2M = r_\sg(t) + c_{12}(t)\sqrt{x} + \mathcal{O}(x), \qquad h = -\frac{1}{2}\ln{\frac{x}{\xi(t)}} + \mathcal{O}\big(\sqrt{x}\big),
\ee
where $\Upsilon$ and $\xi$ are functions of time, $x \defeq r - r_\sg(t)$, $c_{12}(t)= -4\sqrt{\pi}r_\sg^{3/2}\Upsilon$, and the three parameters satisfy
\be
r_\sg' = \mp 4\sqrt{\pi r_\sg \xi}\Upsilon.
\ee
The Schwarzschild radius evolves as a timelike hypersurface. The interpretation of the solutions becomes more transparent when using null coordinates. For $r_\sg' < 0$, the $(v,r)$ coordinates are regular across the horizon and the solution describes an evaporating black hole. Conversely, $r_\sg' > 0$ corresponds to an expanding white hole with regular $(u,r)$ coordinate description. Notably, static solutions are not feasible in the $k = 0$ class.

The case $k = 1$ is more involved. The metric functions of the allowed dynamical solutions are
\be
2M = r + c_{32}(t)x^{3/2} + \mathcal{O}(x^2), \qquad h = -\frac{3}{2}\ln{\frac{x}{\xi(t)}} + \mathcal{O}\big(\sqrt{x}\big),
\ee
where for our purposes the only important property is that $c_{32} < 0$. As before, the solution describes a black or white hole depending on the sign of $r_\sg'$.

In $(v,r)$ coordinates, black holes of both classes are described by
\be
2M_+(v,r) = r_+(v) + w_1(v)y + \mathcal{O}(y^2), \qquad h_+(v,r) = h_1(v)y + \mathcal{O}(y^2), \label{kV}
\ee
where $y \defeq r - r_+(v)$. We note that $w_1 \leqslant 1$, with equality if and only if the solution belongs to the $k = 1$ class.  Due to the invariance of the MSH mass, $M_+(v,r) = M(t(v,r), r)$, it follows that $r_+(v) = r_\sg(t(v, r_+))$. Although the apparent horizon is generally observer-dependent, it remains invariant under all foliations that respect spherical symmetry. Hence, for a \pbh, it is possible to unambiguously define certain events; one example is the instant of formation.  At that moment, the \pbh\ is described by a $k = 1$ solution, which then transitions smoothly into a $k = 0$ solution \cite{MMT:22,MS:23}.

\begin{table}[!htbp]
\footnotesize
\def\arraystretch{2.8}
\centering
\resizebox{0.99\textwidth}{!}{
\begin{tabular}{|>{\centering\arraybackslash}m{0.1\linewidth}|>{\centering\arraybackslash}m{0.15\linewidth}|>{\centering\arraybackslash}m{0.15\linewidth}|>{\centering\arraybackslash}m{0.09\linewidth}|>{\centering\arraybackslash}m{0.15\linewidth}|>{\centering\arraybackslash}m{0.05\linewidth}||>{\centering\arraybackslash}m{0.04\linewidth}|>{\centering\arraybackslash}m{0.04\linewidth}||>{\centering\arraybackslash}m{0.04\linewidth}|>{\centering\arraybackslash}m{0.04\linewidth}||>{\centering\arraybackslash}m{0.04\linewidth}|}
\hline
\raisebox{0.5ex}{Type} & \raisebox{0.5ex}{$\rho$}  & \raisebox{0.5ex}{$p$} & \raisebox{0.5ex}{$p_\|$} & \raisebox{0.5ex}{$\phi$} & \raisebox{0.4ex}{$\mathrm{sgn}\,\Delta$} & \raisebox{0.4ex}{I} & \raisebox{0.4ex}{II} & \raisebox{0.4ex}{$B_1$} & \raisebox{0.4ex}{$B_2$} & \raisebox{0.4ex}{A} \\
\hline\hline
\raisebox{0.6ex}{$k = 0$} & \raisebox{0.6ex}{$\displaystyle\frac{c_{12}(t)}{16\pi r_g(t)^2 \sqrt{x}}$} & \raisebox{0.6ex}{$\displaystyle \frac{c_{12}(t)}{16 \pi r_\sg(t)^2 \sqrt{x}}$} & \raisebox{0.6ex}{finite} & \raisebox{0.6ex}{$\displaystyle\frac{|c_{12}(t)|}{16 \pi r_\sg(t)^2 \sqrt{x}}$} & \raisebox{0.5ex}{$-$} & \raisebox{0.5ex}{$\xmark$} & \raisebox{0.5ex}{$\xmark$}  & \raisebox{0.5ex}{$\xmark$} & \raisebox{0.5ex}{$\xmark$}  & \raisebox{0.5ex}{$\textbf{?}$} \\
\hline
\raisebox{0.6ex}{$k = 1$} & \raisebox{0.6ex}{$\displaystyle \frac{1}{8 \pi r_\sg(t)^2}$} & \raisebox{0.6ex}{$\displaystyle -\frac{1}{8 \pi r_\sg(t)^2}$} & \raisebox{0.6ex}{finite} & \raisebox{0.5ex}{$\displaystyle \frac{3 |c_{32}(t)r'_\sg(t)|x^2}{32 \pi r_\sg(t) }$} & \raisebox{0.5ex}{\textbf{?}} & \raisebox{0.5ex}{$\xmark$} & \raisebox{0.5ex}{$\xmark$} & \raisebox{0.5ex}{\textbf{?}} & \raisebox{0.5ex}{$\cmark$} & \raisebox{0.5ex}{$\cmark$} \\
\hline
\end{tabular}}
\caption{Overview of the leading terms of the near-horizon expansions of energy density $\rho$, radial pressure $p$, tangential pressure $p_\|$, flux $\phi$, and pressure anisotropy $\Delta\defeq p - p_\|$ for the two allowed classes of solutions. The conditions on the EMT are presented, with $(\cmark)$ indicating satisfaction, $(\xmark)$ denoting violation at the horizon, and $(\textbf{?})$ indicating that the answer depends on higher-order terms.}
\label{table:EMT}
\end{table}

Having a finite time of horizon crossing according to a distant clock requires that $\lim_{r \to r_\sg} e^h f$ is finite. Hence, no static ($k = 1$) solution satisfies this requirement. The \pbh~ solutions have many interesting properties, but we restrict the discussion only to those that are directly relevant to the conditions of the Buchdahl theorem \cite{MMT:22}.

The $k = 0$ and $k = 1$ solutions violate the null energy condition \cite{HE:73,exact:03}. In both cases, the pressure is negative, and for $k = 0$, the energy density is also negative. As a result, condition I is always violated.

Moreover, since $p \neq p_\|$ in these solutions, describing the EMT in the dynamical cases requires \textit{at least} two fluids. This is most conveniently represented in the Schwarzschild frame as
\begin{align}
T_{\hat{\mu} \hat{\nu}} =
\left(
\begin{array}{@{}r@{\quad}r@{\quad}r@{\quad}r@{}}
\rho & -\phi & 0 & 0 \\
-\phi & p & 0 & 0 \\
0 & 0 & p_\parallel & 0 \\
0 & 0 & 0 & p_\parallel
\end{array}
\right)
=
\left(
\begin{array}{@{}r@{\quad}r@{\hspace*{0.7cm}}r@{\hspace*{0.7cm}}r@{}}
\phi & -\phi & 0 & 0 \\
-\phi & \phi & 0 & 0 \\
0 & 0 & 0 & 0 \\
0 & 0 & 0 & 0
\end{array}
\right)
+
\left(
\begin{array}{@{}r@{\hspace*{0.5cm}}r@{\quad}r@{\quad}r@{}}
\tilde{\rho} & 0 & 0 & 0 \\
0 & \tilde{p} & 0 & 0 \\
0 & 0 & p_\parallel & 0 \\
0 & 0 & 0 & p_\parallel
\end{array}
\right),
\end{align}
with $\tilde{\rho} \defeq \rho - \phi$ and $\tilde{p} \defeq p - \phi$. This allows the EMT to be decomposed into a radially moving null dust and an anisotropic static fluid with time-dependent parameters. For $k = 0$, the null fluid density, radial pressure, and flux diverge in the Schwarzschild frame but are finite in a frame comoving with the infalling observer.

The near-horizon expansion yields an anisotropy $\Delta = p - p_\| < 0$ for $k = 0$ solutions ($p$ is divergent and negative, $p_\|$ is finite). Hence, condition $B_1$ is violated. For $k = 1$ solutions, the sign of $\Delta$ depends on higher-order terms.

   As can be readily inferred from Table~\ref{table:EMT} the derivative $\pad_r \rho > 0$ near the horizon for $k = 0$, thereby violating condition $B_2$. In contrast, for dynamic $k = 1$ solutions,
\begin{align}
\frac{\partial\rho}{\partial r} = \frac{3 c_{32}(t)}{32 \pi r_\sg(t)^2 \sqrt{x}} + \mathcal{O}\big(x^0\big) < 0\ .
\end{align}
Finally, condition A is trivially satisfied at the apparent horizon by all admissible solutions. In its vicinity, it holds for $k = 1$ (albeit only momentarily), while for $k = 0$ the outcome depends on higher-order terms.
\vspace*{-0.6cm}

\section{Conclusions}

Physical black holes — trapped regions of spacetime whose outer apparent horizon is accessible to distant observers — are exotic by construction. Indeed, a theorem by Hawking and Ellis \cite{HE:73,FN:98} requires violation of the null energy condition for their visibility from away. Our constructions of $\Phi$BHs provide explicit demonstration of this result: attempts to enforce the NEC yield complex-valued solutions \cite{BMMT:19}. These configurations also exhibit strong anisotropy. In fact, after formation, \pbh{}s require a null fluid component in their EMT and violate all four conditions of the Buchdahl theorem.

The key insight is that the requirement of a horizon is far from innocent. \pbh{}s are significantly more exotic than horizonless ECOs. A sustained violation of the NEC is often seen as a sign that semiclassical physics breaks down \cite{HV:20} and that quantum gravitational effects become important at the horizon scale \cite{MMT:22, A:20}. Thus, one possibility is to reject all exoticism as unacceptable. In that case, only scenario (i), i.e., asymptotic collapse, remains — and the event horizon, the information loss problem, and the rich inner life of black holes are excluded from physics. Alternatively, all exotic objects — with or without horizons — are viable contenders to describe the ABHs. Scenarios (ii) and (iii), if realised, promise genuinely new physics. Only   observations will tell.

	\acknowledgments
	%We would like to thank ... for useful discussions and helpful comments.
	IS is supported by an International Macquarie University Research Excellence Scholarship. The work of DRT is supported by the ARC Discovery project Grant No. DP210101279 and the Schwinger Foundation.

\end{document}